\documentclass[aps,prl,preprint,superscriptaddress]{revtex4}
\usepackage{graphicx}% Include figure files

\begin{document}

%\preprint{}

\title{On the NuTeV anomaly and the asymmetry of the strange sea\\
in the nucleon}


\author{M.~Wakamatsu}
%\email[]{wakamatu@phys.sci.osaka-u.ac.jp}
%\thanks{}
\affiliation{Department of Physics, Faculty of Science, \\
Osaka University, \\
Toyonaka, Osaka 560-0043, JAPAN}


%\date{\today}

\begin{abstract}
\ \ Our recent theoretical analysis based on the flavor SU(3) chiral quark
soliton model predicts fairly large particle-antiparticle asymmetry of
the strange sea in the nucleon. We point out that the predicted magnitude
of asymmetry is large enough to solely resolve the so-called NuTeV
anomaly on the fundamental parameter $\sin^2 \theta_W$ in the standard
model.
\end{abstract}

%\keywords{}

\maketitle

%\section{}
%\subsection{}
%\subsubsection{}

At the early stage of high-energy deep-inelastic scattering analyses, 
it was a common assumption that the sea quark distributions in the nucleon
are flavor SU(3) symmetric.  It is clear by now, however, that there is no
sound theoretical reason to justify this dogma. In fact, the isospin SU(2)
asymmetry of the nucleon sea, i.e. the asymmetry of $\bar{u}$ and
$\bar{d}$ distributions in the proton has been definitely established by
the NMC measurement \cite{NMC91}\nocite{HM90}\nocite{KK91}--\cite{Waka92}.
Similarly, it is highly probable that the 
momentum distributions of strange quarks and antiquarks are not the same
despite the constraint that the total numbers of $s$- and
$\bar{s}$-quarks are precisely equal in the
nucleon \cite{ST87}\nocite{BW92}\nocite{BM96}\nocite{CS2003}
\nocite{DM2004}-\cite{AI2004}.
The possible asymmetry of the $s$- and $\bar{s}$-quark distributions in
the nucleon has attracted a renewed interest after it was recognized that 
it plays a crucial rule in the interpretation of NuTeV determination
of the weak mixing angle 
\cite{NuTeV2002A}\nocite{NuTeV2002B}\nocite{Davidson2002}
\nocite{Olness2003}\nocite{Kretzer2004A}\nocite{Kretzer2004B}
--\cite{Londergan2004}.
The NuTeV Collaboration extracted the value of 
$\sin^2 \theta_W$ by measuring the ratio of neutrino neutral-current and 
charged-current cross sections on iron.
The measured ratio $R^-$ (the so-called Paschos-Wolfenstein
ratio \cite{PW73}) is related to the Weinberg
angle $\theta_W$ by \cite{Davidson2002}\nocite{Olness2003}
\nocite{Kretzer2004A}--\cite{Kretzer2004B}
\begin{eqnarray}
 R^- &\equiv& \frac{\sigma_{NC}^{\nu} - \sigma_{NC}^{\bar{\nu}}}
 {\sigma_{CC}^{\nu} - \sigma_{CC}^{\bar{\nu}}} \nonumber \\
 &=& \frac{1}{2} - \sin^2 \theta_W + \delta R_A^- + \delta R_{QCD}^- 
 + \delta R_{E W}^- ,
\end{eqnarray}
where the three correction terms respectively stand for the target 
non-isoscalarity correction ($\delta R_A^-$), QCD corrections
($\delta R_{QCD}^-$) and higher-order electroweak corrections
($\delta R_{E W}^-$).
The QCD corrections come from three main sources as 
\begin{equation}
 \delta R_{QCD}^- = \delta R_s^- + \delta R_I^- + \delta R_{NLO}^- ,
\end{equation}
where $\delta R_s^-, \delta R_I^-$ and $\delta R_{NLO}^-$ respectively 
stand for possible strange-sea asymmetry ($s^- \equiv s - \bar{s} \neq 0$), 
isospin violation ($u_{p,n} \neq d_{n,p}$) effects in the parton density 
of the nucleon, and the NLO corrections. In  the present study, we focus
on the first correction due to the possible asymmetry of the strange sea
in the nucleon. Approximately, it is given by \cite{Davidson2002}
\begin{equation}
 \delta R_s^- \simeq - \,\left(\frac{1}{2} - 
 \frac{7}{6} \sin^2 \theta_W \right) \, \frac{[S^-]}{[Q^-]} ,\label{ratio}
\end{equation}
where 
\begin{eqnarray}
 \, [S^-] &\equiv& \int_0^1 x \,[ s(x) - \bar{s}(x) ] d x ,\label{sm} \\
 \, [Q^-] &\equiv& \int_0^1 x \,[ u(x) - \bar{u}(x) + d(x) - \bar{d}(x) ] 
 d x . \label{qm}
\end{eqnarray}
Recently, the CTEQ group performed a global PDF fit including the NuTeV
``dimuon events" on the neutrino and antineutrino-production of
charm \cite{Olness2003,Kretzer2004A}.
Their analysis leads to a central value $[S^-] \simeq 0.002$ and
conservative bound
\begin{equation}
 - 0.001 < [S^-] < 0.004 .
\end{equation}
Note that the positive moment $[S^-]$, which means that the momentum
distributions of the $s$-quark is harder than that of the $\bar{s}$-quark
in the nucleon, works to reduce the discrepancy between the NuTeV
determination of $\sin^2 \theta_W$ \cite{NuTeV2002A,NuTeV2002B} and
the world average of other measurements \cite{Abbaneo2001}.

Since the distribution of sea quarks and antiquarks generated through the
perturbative QCD evolution are necessarily CP symmetric, the cause of 
asymmetry of the nucleon strange sea must be of nonperturbative origin.
(Note however the recent claim that the three loop QCD correction may
generate a sizable strange-quark asymmetry in the
nucleon \cite{CFRV2004}.)
As discussed by many authors, the most plausible source of the asymmetry
may be the virtual fluctuation process of the physical
proton into the $\Lambda K^+$ intermediate
state \cite{ST87}\nocite{BW92}\nocite{BM96}\nocite{CS2003}
\nocite{DM2004}--\cite{AI2004}.
Since the $s$- and $\bar{s}$-quarks in the intermediate state are
contained in totally different type of hadrons, i.e. a baryon and a meson,
their helicity and momentum distribution can be significantly different.
It was argued that this ``kaon-cloud picture" of the nucleon leads to
several interesting predictions, such as

\begin{description}
\item{(1)} $s$-quarks carry more momentum fraction, than $\bar{s}$-quarks,
i.e. $\int_0^1 x s(x) dx > \int_0^1 x \bar{s}(x) dx$. 
\item{(2)} $s$-quarks are polarized antiparallel to the initial proton
spin.
\end{description}

Although intuitively very appealing (and we believe it contains a piece of 
the truth), the predictions of the kaon-cloud model should be taken as 
only suggestive. The reason is clear. For obtaining the desired strange 
and antistrange distributions of the nucleon in this model, one need two 
basic quantities which are not known very well. The one is the so-called 
meson-cloud-model {\it fluctuation functions}, which give the probability
to find the baryon or meson with some longitudinal momentum fraction.
The other is the strange valence parton distributions of the
{\it constituents of cloud}, i.e. $\Lambda$ and $K^+$.
Even worse, it is far from clear how many mesons, 
besides the pseudoscalar octet, one should take into account. In fact,
in a recent paper, Cao and Signal estimated the strange sea distributions
by taking account not only of $p \rightarrow \Lambda K^+$ fluctuation, but 
also of some other fluctuations into $\Lambda K^{* -}, \Sigma K$ and 
$\Sigma K$ intermediate states \cite{CS2003}. Embarrassingly, their
conclusion was that the $s$-$\bar{s}$ momentum asymmetry generated
by the $\Lambda K^+$ fluctuation is cancelled nearly completely by the
fluctuation containing $K^*$ clouds.

Obviously, what we need is more systematic approach, which is free from
the above-mentioned theoretical ambiguities.
We claim that chiral quark soliton model (CQSM) is the best candidate
to meet such requirements. The great advantage 
of this effective quark model is that the Goldstone bosons appear
(automatically) as composites \cite{DPP88,WY91},
which enables us to introduce effects of
meson cloud without worrying about many ambiguities and
complexities inherent in the meson-cloud convolution model.
It has already been shown that, without introducing any adjustable
parameter except for the initial-energy scale of the $Q^2$-evolution,
the CQSM can explain nearly all the qualitatively noticeable features
of the recent high-energy deep-inelastic scattering
observables \cite{DPPPW96}\nocite{DPPPW97}\nocite{WGR96}\nocite{WGR97}
\nocite{WK98}\nocite{Poby99}\nocite{WK99}
\nocite{Waka92P}-\cite{WW2000}.
It naturally explains the excess of $\bar{d}$-sea over the $\bar{u}$-sea
in the proton \cite{Waka92,WK98,Poby99}.
It also reproduces the characteristic
features of the observed longitudinally polarized structure functions for 
the proton, the neutron and the deuteron \cite{WK99}. The most puzzling
observation, i.e. unexpectedly small quark spin fraction of the nucleon,
can also be explained in no need of large gluon polarization at the low 
renormalization scale \cite{WY91,Waka92P,WW2000}.
Recently, we have also addressed the problem of quark-antiquark asymmetry
of the nucleon strange sea distribution based on the CQSM generalized to
flavor SU(3) \cite{Waka2003A,Waka2003B}. (See also \cite{SRW99}.)
It turned out that the predictions of the SU(3) CQSM supports
the general idea of the most naive 
kaon cloud model at least qualitatively.
It predicts a sizable amount of $s$-$\bar{s}$ momentum asymmetry
in such a way that $\int_0^1 x s(x) d x > \int_0^1 x \bar{s}(x) dx$.
It also predicts that the $s$-quark is negatively polarized with respect
to the proton spin direction, while the polarization of $\bar{s}$-quark
is relatively small. In consideration of the impact of the NuTeV anomaly, 
here we reexamine the problem of the $s$-$\bar{s}$ momentum asymmetry 
and see what this unique model can say about it.

The basic Lagrangian of the SU(3) CQSM is given as 
\begin{equation}
 {\cal L} = \bar{\psi} (x) (i\not\!\partial - M U^{\gamma_5}(x) 
 - \Delta m_s P_s ) \psi (x) ,
\end{equation}
with
\begin{equation}
 U^{\gamma_5} (x) = e^{i \gamma_5 \pi (x) / f_{\pi}}, \ \ \ 
 \pi (x) = \sum_{a = 1}^8 \pi_a (x) \lambda_a ,
\end{equation}
and 
\begin{equation}
 \Delta m_s P_s = \Delta m_s \,
 \left(\frac{1}{3} - \frac{1}{\sqrt{3}} \lambda_8 \right)
 =
 \left(
 \begin{array}{ccc}
 0 & 0 & 0 \\
 0 & 0 & 0\\
 0 & 0 & \Delta m_s
 \end{array}
 \right) .
\end{equation}
It is a straightforward generalization of the corresponding SU(2) model,
except for one important new feature, i.e. the presence of the sizably
large SU(3) symmetry breaking term arising from the effective mass 
difference $\Delta m_s$ between the strange and nonstrange quarks. 
Since the dynamical quark mass $M$ is already fixed to the value
$M \simeq 375 \,\mbox{MeV}$ from the phenomenology of nucleon low energy
observables within the SU(2) model, this mass difference $\Delta m_s$
is the {\it only one additional parameter} in the flavor SU(3)
generalization of the CQSM.
Since the detail of the model was already explained
in \cite{Waka2003A,Waka2003B}, here we only 
want to emphasize that the SU(3) symmetry breaking effects can be 
estimated by using the first order perturbation theory in the parameter 
$\Delta m_s$. We believe this treatment is justified (at least partially),
since the effective mass difference $\Delta m_s$ of the order of 100 MeV
is much smaller than the typical energy scale of baryons.
Naturally, the most sensitive quantities to the parameter $\Delta m_s$
are magnitudes of the $s$- and $\bar{s}$-quark distributions.
In \cite{Waka2003A}, we determined
$\Delta m_s$ so as to reproduce the CCFR data for $s(x)$ and 
$\bar{s}(x)$ distribution, which was extracted under the constraint
$s(x) = \bar{s}(x)$. The overall success of the theory is obtained with
the value of $\Delta m_s$ around 100 MeV.

\begin{figure}[htb] \centering
\begin{center}
 \includegraphics[width=9.0cm,height=7.5cm]{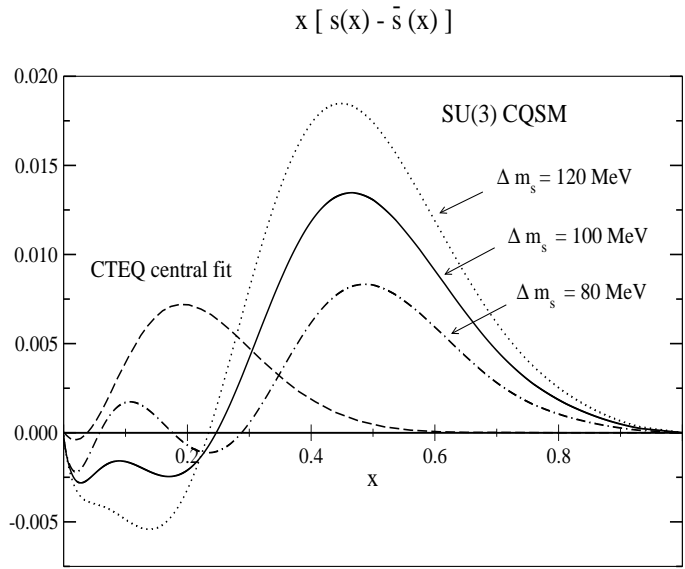}
\end{center}
\vspace*{-0.5cm}
\renewcommand{\baselinestretch}{1.20}
\caption{The theoretical predictions of the SU(3) CQSM for
$ x [ s(x) - \bar{s} (x) ]$ evolved to $Q^2 = 16 \,\mbox{GeV}^2$ with
$\Delta m_s = 80 \,\mbox{MeV}$, $100 \,\mbox{MeV}$ and $120 \,
\mbox{MeV}$ in comparison with the CTEQ central fit.}
\label{fig:ssbasymm}
\end{figure}%

As discussed in \cite{Waka2003A}, the difference
function $s(x) - \bar{s}(x)$ is also very sensitive to the SU(3)
symmetry breaking effect.
Here we examine the effect of the variation of the parameter
$\Delta m_s$ around its central value above to see its influence on the
NuTeV analysis.  Shown in Fig.1 are the theoretical predictions of the
SU(3) CQSM for $x [ s(x) - \bar{s}(x) ]$. Here the dash-dotted, the solid
and the dotted curves are obtained with $\Delta m_s = 80 \,\mbox{MeV}$,
100 MeV and 120 MeV, respectively. Also shown for comparison is the
central fit of the CTEQ group (dashed curve).
In view of the sizably large uncertainties coming from 
using the neutrino data on nuclear target, the qualitative agreement 
between the theory and CTEQ fit seems encouraging.
One clearly sees a common tendency in the results of the two analyses,
which are of totally different nature. Both shows that the momentum 
distributions of the $s$-quark is harder than that of the $\bar{s}$-quark.
Now the question is the size or the magnitude of the predicted momentum 
asymmetry.

\vspace{3mm}
\newcommand{\lw}[1]{\smash{\lower2.ex\hbox{#1}}}
\begin{table}[h]
\caption{The prediction of the SU(3) CQSM for the second moments
$[S^-], [Q^-]$ and the correction $\delta R_s^-$ to the
Paschos-Wolfenstwein ratio $R^-$ in dependence of $\Delta m_s$.}
\vspace{2mm}
\begin{center}
\renewcommand{\arraystretch}{1.0}
\begin{tabular}{|c|c|c|c|}
\hline\hline
\ \ $\Delta m_s \,(\mbox{MeV})$ \ \ & 80 & 100 & 120 \\
\hline\hline
$S^-$ & \ \ 0.0025 \ \ & \ \ 0.0040 \ \ & \ \ 0.0055 \ \ \\
\hline
$Q^-$ & \ \ 0.226 \ \ & \ \ 0.227 \ \ & 0.228 \ \ \\
\hline
$\delta R_s^-$ & \ \ - \,0.0034 \ \ & \ \ - \,0.0055 \ \ & 
\ \ - \,0.0075 \ \ \\
\hline\hline
\end{tabular}
\end{center}
\end{table}

The crucial quantities here are the second moments $[S^-]$ and $[Q^-]$ 
defined in (\ref{sm}),(\ref{qm}), which in turn give $\delta R_s^-$
according to (\ref{ratio}). 
We show in table 1 the predictions of the SU(3) CQSM for
$[S^-]$, $[Q^-]$ and $\delta R_s^-$ obtained with the three choices
of the parameter $\Delta m_s$, i.e. $\Delta m_s = 80 \,\mbox{MeV}$,
$100 \,\mbox{MeV}$, and $120 \,\mbox{MeV}$.
The CQSM prediction for $[S^-]$ ranges from 0.0025 to 0.0055 with
the central value
\begin{equation}
 [S^-]_{CQSM} = +0.004 ,
\end{equation}
corresponding to the above choices of $\Delta m_s$. This prediction may
be compared with the central value
\begin{equation}
 [S^-]_{CTEQ} = +0.002 ,
\end{equation}
and the conservative bounds
\begin{equation}
 -0.001 < [S^-]_{CTEQ} < +0.004 ,
\end{equation}
obtained by the CTEQ PDF fit. Translating into a shift of the Weinberg
angle, the SU(3) CQSM gives
\begin{equation}
 - 0.0075 < \delta (\sin^2 \theta_W)_{CQSM} < - 0.0034 ,
\end{equation}
with the central value $\delta (\sin^2 \theta_W)_{CQSM} = -0.0055$.
Note that the central value obtained with $\Delta m_s = 100 \,\mbox{MeV}$
is large enough to fill a gap between the NuTeV determination of
$\sin^2 \theta_W$ ($0.2277 \pm 0.0013 \pm 0.0009$) and the world
average of other measurements ($0.2227 \pm 0.0004$). Even the
conservative estimate obtained with a smaller value of
$\Delta m_s = 80 \,\mbox{MeV}$ explains nearly $70 \,\%$ of the discrepancy.

To summarize, before extracting any new physics beyond the standard model
from the NuTeV anomaly, one should first worry about theoretical 
uncertainties mainly due to QCD. Undoubtedly, the possible asymmetry of
the strange sea in the nucleon is one of the most important factors
that we must take seriously. In fact, the existence of the asymmetry
seems an unavoidable physical consequence of chiral symmetry of QCD.
The widely-accepted scenario of spontaneous breakdown of this symmetry
dictates the appearance of low mass pseudoscalar octet, which in turn
make these mesons the source of the energetically lowest excitation of
nucleon with intrinsically generated sea quarks.
This scenario, which is completely consistent with the well-established 
dominance of the $\bar{d}$-sea over the $\bar{u}$-sea in the proton,
also indicates the asymmetry of nucleon strange sea.
Unfortunately, the simplest candidate, i.e. the meson-cloud convolution
model, which realizes this idea in the most direct way, suffers from
several theoretical uncertainties and its predictions are now widely
diverse.
This is to be contrasted with the flavor SU(3) generalization of the
CQSM. Only one parameter of this model is the effective mass difference
between the strange and nonstrange quarks, which gives the measure of
the SU(3) symmetry breaking. Within the reasonable range of this
parameter, the SU(3) CQSM has been shown to predict a sizable amount of
asymmetry of the $s$- and $\bar{s}$-quark distributions in the nucleon,
which is large enough to resolve the NuTeV anomaly.
In any case, one important fact has become apparent through the
investigations inspired by the NuTeV report. The neutrino-induced
DIS measurement has become close to practical use as a tool to
probe the internal structure of the nucleon. We expect that neutrino DIS
scattering experiments to be carried out in the near future will enable
direct and more accurate determination of the light-flavor
sea-quark distribution functions in the nucleon.

\vspace{3mm}
\begin{acknowledgments}
This work is supported in part by a Grant-in-Aid for Scientific
Research for Ministry of Education, Culture, Sports, Science
and Technology, Japan (No.~C-16540253)
\end{acknowledgments}



\begin{thebibliography}{10}

\bibitem{NMC91}
NMC Collaboration~: P.~Amaudruz~et al.
\newblock {\em Phys. Rev. Lett.}, 66:2712, 1991.

\bibitem{HM90}
E.M Henley and G.A. Miller.
\newblock {\em Phys. Lett.}, B251:453, 1990.

\bibitem{KK91}
S.~Kumano and J.T. Londergan.
\newblock {\em Phys. Rev.}, D43:59, 1991.

\bibitem{Waka92}
M.~Wakamatsu.
\newblock {\em Phys. Rev.}, D46:3762, 1992.

\bibitem{ST87}
A.I. Signal and A.W. Thomas.
\newblock {\em Phys. Lett.}, B191:205, 1987.

\bibitem{BW92}
M.~Burkardt and B.J. Warr.
\newblock {\em Phys. Rev.}, D45:958, 1992.

\bibitem{BM96}
S.J. Brodsky and B.-Q. Ma.
\newblock {\em Phys. Lett.}, B381:317, 1996.

\bibitem{CS2003}
F.-G. Cao and A.-I. Signal.
\newblock {\em Phys. Lett.}, B559:229, 2003.

\bibitem{DM2004}
Y.~Ding and B.-Q. Ma.
\newblock {\em Phys. Lett.}, B5901:216, 2004.

\bibitem{AI2004}
J.~Alwall and G.~Ingelman.
\newblock {\em hep-ph/0407364}.

\bibitem{NuTeV2002A}
NuTeV Collaboration~: G.P.~Zeller~et al.
\newblock {\em Phys. Rev. Lett.}, 88:091802, 2002.

\bibitem{NuTeV2002B}
NuTeV Collaboration~: G.P.~Zeller~et al.
\newblock {\em Phys. Rev.}, D65:111103, 2002.

\bibitem{Davidson2002}
S.~Davidson, S.~Forte, P.~Gambino, N.~Rius, and A.~Strumia.
\newblock {\em JHEP}, 02:037, 2002.

\bibitem{Olness2003}
F.~Olness, J.~Pumplin, D.~Stump, J.~Huston, P.~Nadolsky, H.-L. Lai, S.~Kretzer,
  J.F. Owens, and W.K. Tung.
\newblock {\em hep-ph/0312323}.

\bibitem{Kretzer2004A}
S.~Kretzer, F.~Olness, J.~Pumplin, D.~Stump, W.K. Tung, and M.H. Reno.
\newblock {\em Phys. Rev. Lett.}, 93:041802, 2004.

\bibitem{Kretzer2004B}
S.~Kretzer.
\newblock {\em hep-ph/0405221}.

\bibitem{Londergan2004}
J.T. Londergan.
\newblock {\em hep-ph/0408243}.

\bibitem{PW73}
E.A.. Paschos and L.~Wolfenstein.
\newblock {\em Phys. Rev.}, D7:91, 1973.

\bibitem{Abbaneo2001}
D.~Abbaneo et~al.
\newblock {\em CERN-EP/2001-98, hep-ex/0112021}, 2001.

\bibitem{CFRV2004}
S.~Catani, D.~Florian, G.~Rodrigo, and W.~Vogelsang.
\newblock {\em Phys. Rev. Lett.}, 93:152003, 2004.

\bibitem{DPP88}
D.I. Diakonov, V.Yu. Petrov, and P.V. Pobylitsa.
\newblock {\em Nucl. Phys.}, B306:809, 1988.

\bibitem{WY91}
M.~Wakamatsu and H.~Yoshiki.
\newblock {\em Nucl. Phys.}, A524:561, 1991.

\bibitem{DPPPW96}
D.I. Diakonov, V.Yu. Petrov, P.V. Pobylitsa, M.V. Polyakov, and C.~Weiss.
\newblock {\em Nucl. Phys.}, B480:341, 1996.

\bibitem{DPPPW97}
D.I. Diakonov, V.Yu. Petrov, P.V. Pobylitsa, M.V. Polyakov, and C.~Weiss.
\newblock {\em Phys. Rev.}, D56:4069, 1997.

\bibitem{WGR96}
H.~Weigel, L.P. Gamberg, and H.~Reinhardt.
\newblock {\em Mod. Phys. Lett.}, A11:3021, 1996.

\bibitem{WGR97}
H.~Weigel, L.P. Gamberg, and H.~Reinhardt.
\newblock {\em Phys. Rev.}, D55:6910, 1997.

\bibitem{WK98}
M.~Wakamatsu and T.~Kubota.
\newblock {\em Phys. Rev.}, D56:4069, 1998.

\bibitem{Poby99}
P.V. Pobylitsa, M.V. Polyakov, K.~Goeke, T.~Watabe, and C.~Weiss.
\newblock {\em Phys. Rev.}, D59:034024, 1999.

\bibitem{WK99}
M.~Wakamatsu and T.~Kubota.
\newblock {\em Phys. Rev.}, D60:034020, 1999.

\bibitem{Waka92P}
M.~Wakamatsu.
\newblock {\em Prog. Theor. Phys. Suppl.}, 109:115, 1992.

\bibitem{WW2000}
M.~Wakamatsu and T.~Watabe.
\newblock {\em Phys. Rev.}, D62:054009, 2000.

\bibitem{Waka2003A}
M.~Wakamatsu.
\newblock {\em Phys. Rev.}, D67:034005, 2003.

\bibitem{Waka2003B}
M.~Wakamatsu.
\newblock {\em Phys. Rev.}, D67:034006, 2003.

\bibitem{SRW99}
O.~Schroeder, H.~Reinhardt, and H.~Weigel.
\newblock {\em Nucl. Phys.}, A651:174, 1999.

\end{thebibliography}
\end{document}